\documentclass[11pt]{article}
\usepackage{graphicx}
\usepackage{amssymb}
\usepackage{amsmath}
\usepackage{graphicx}
\usepackage{amstext}
\usepackage{esint}
\usepackage[utf8]{inputenc}
\usepackage{color}
\usepackage{cite}

\def\beq{\begin{eqnarray}}    
\def\eeq{\end{eqnarray}}      

\newcommand{\rL}{\rho_\Lambda}

\newcommand{\CC}{\Lambda}

\newcommand{\OMo}{\Omega_{m}^0}
\newcommand{\ORo}{\Omega_{r}^0}

\newcommand{\OLo}{\Omega^0_{\Lambda}}

\newcommand{\rco}{\rho^0_{c}}

\newcommand{\rLo}{\rho_{\CC}^0}

\newcommand{\rDE}{\rho_{\rm DE}}

\newcommand{\oD}{\omega_{\rm BD}}
\newcommand{\newtext}[1]{{\textcolor{black}{#1}}}

\newcommand{\dpsi}{\dot{\psi}}
\newcommand{\ddpsi}{\ddot{\psi}}
\newcommand{\wBD}{\omega_{\rm BD}}

\newcommand{\fracdpsipsi}{\frac{\dpsi}{\psi}}
\newcommand{\fracddpsipsi}{\frac{\ddpsi}{\psi}}

\begin{document}



 \hyphenation{nu-cleo-syn-the-sis u-sing si-mu-la-te ma-king
cos-mo-lo-gy know-led-ge e-vi-den-ce stu-dies be-ha-vi-or
res-pec-ti-ve-ly appro-xi-ma-te-ly gra-vi-ty sca-ling
ge-ne-ra-li-zed re-mai-ning}




\begin{center}
{\it\LARGE Brans-Dicke cosmology mimicking running vacuum} \vskip 2mm

 \vskip 8mm

\textbf{\large Javier de Cruz P\'erez and Joan Sol\`a Peracaula}

\vskip 0.5cm
Departament de F\'isica Qu\`antica i Astrof\'isica, and Institute of Cosmos Sciences,\\ Universitat de Barcelona, \\
Av. Diagonal 647, E-08028 Barcelona, Catalonia, Spain

\vskip0.5cm

\vskip0.4cm

E-mails:   decruz@fqa.ub.edu, sola@fqa.ub.edu

 \vskip2mm

\end{center}
\vskip 15mm

\begin{quotation}
\noindent {\large\it \underline{Abstract}}.
Brans-Dicke (BD) cosmology is reconsidered from an approach in which the model can be formulated in $\Lambda$CDM form, but at the expense of replacing the rigid cosmological constant $\Lambda$   with a dynamical quasivacuum  component which depends on  a small parameter $\epsilon$. The product of $\epsilon$ times the BD-parameter $\omega_{\rm BD}$ becomes exactly determined in terms of the ordinary cosmological parameters. The GR limit is recovered for $\omega_{\rm BD}\to\infty$ when $\epsilon\to 0$.  We solve the background cosmology of the model as well as the perturbations equations. When fitted to the cosmological data, we find that the BD-cosmology, transcribed in such an effective GR form, appears to be competitive with the $\Lambda$CDM and emulates the running vacuum model.
\end{quotation}
\vskip 5mm

\begin{quotation}
\end{quotation}

\newpage


\newpage


\section{Introduction}\label{intro}

For slightly more than 20 years we know with a substantial degree of certainty that the universe is in accelerated expansion\,\cite{SNIaRiess,SNIaPerl}. Such knowledge is, however, a pure kinematical result based on assuming General Relativity (GR). However, it does not mean that we really understand the primary cause for such an acceleration and the ultimate cosmological context where it has been measured (see below).  The canonical picture in the framework of GR is to assume that it is caused by a rigid cosmological constant (CC) term, $\CC$, in Einstein's equations, whose value has by now been pinned down with remarkable precision from a rich asset of observations\,\cite{Planck2018}. Unfortunately we do not know what is the  fundamental origin of $\CC$. It is assumed to take a constant value for the entire cosmic history. Not too surprisingly, such oversimplification must be the reason for having to put up since long (and still in our days) with the Cosmological Constant Problem, namely the appalling discrepancy between the measured value of the vacuum energy density in the cosmos,  $\rLo=\CC/(8\pi G)\sim 10^{-47}$ GeV$^4$ (G being Newton's constant), and the predicted one in quantum field theory (QFT) -- which is many orders of magnitude larger\,\cite{Weinberg2000,JSPRev2013}. Even so, the phenomenological status of a rigid cosmological term is thought to be  robust and unbeaten. In fact, the cosmological  model which is built upon the assumed existence of the $\CC$-term and  of dark matter (DM), together with the assumption that the cosmological metric is that of  a spatially flat  Friedmann-Lema\^{\i}tre-Robertson-Walker (FLRW) universe, is called  the ``concordance'' $\CC$CDM model, i.e. the standard model of cosmology\,\cite{Peebles1993,LiddleLyth2000}. It is fairly consistent with a large body of observations, including the high precision data from the cosmic microwave background (CMB) anisotropies\,\cite{Planck2018}.

In spite of the many virtues of the $\CC$CDM, the longstanding (and unaccounted) constancy and value of $\CC$ and the acute severity of the Cosmological Constant Problem has motivated a variety of alternative explanations for the cosmic acceleration beyond the $\CC$-term. Such theories include quintessence and the like and they go under the generic name of dark energy (DE), see e.g. \,\cite{DEBook,PeeblesRatra2003} and references therein. The idea that the DE could be not just the CC of Einstein's equations but a dynamical variable, or just some appropriate function of the cosmic time, has been explored since long ago and sometimes on purely phenomenological grounds\,\cite{Overduin98}; see particularly\,\cite{CCt}. In all these cases the basic framework is GR despite $\CC$ is not constant but a dynamical scalar field or an explicitly time-evolving quantity.

In the sixties, another significant revolution occurred in the gravity context, in which $G$ was boldly assumed to be a dynamical variable rather than a constant of Nature. This proposal clearly departs from the strict GR context. It actually traces back to early ideas in the thirties on the possibility of a time-evolving gravitational constant $G$ by Milne\,\cite{Milne1935} and the suggestion by Dirac of the large number hypothesis \cite{Dirac1937}, which led him also to propose the time evolution of $G$. Along similar lines, Jordan and Fierz speculated that the fine structure constant $\alpha_{\rm em}$ together with $G$ could be both space and time dependent\,\cite{JordanFierz}. Finally, $G$ was formally associated to the existence of a dynamical scalar field $\psi\sim 1/G$ coupled to the curvature.  Such was the famous gravity formulation originally proposed by Brans and Dicke (``BD'' for short) \cite{BD}, which was the first historical attempt to extend GR to accommodate variations in the Newtonian coupling $G$. Subsequently these ideas were generalized in the form of scalar-tensor theories\,\cite{ScalarTensor}.

In this work, we combine the dynamical character of $G$ in the context of BD-gravity with the idea of dynamical DE. More specifically, we show that if one tries to encapsulate the slow evolution of the BD-field in terms of the current GR paradigm (in which $G$ remains constant), the effective theory that emerges is a variant of the $\CC$CDM framework in which $\rho_\Lambda
$ acquires a time-evolving component and plays the role of an approximate dynamical vacuum energy density. The effective model, therefore, is not exactly the traditional $\CC$CDM, but has additional features. This might be an interesting possibility, given the fact that there are currently significant tensions in the context of the strict $\CC$CDM, for example between CMB measurements of the Hubble parameter $H_0$\,\cite{Planck2018}  and local determinations of it\,\cite{RiessH02016,RiessH02018};  and at the same time we have a no less  stubborn discrepancy between the large scale structure (LSS) formation data and the predicted one by the conventional $\CC$CDM, the so-called $\sigma_8$-tension\,\cite{sigma8tension}. Bearing in mind that these tensions could perhaps be alleviated in the context of dynamical vacuum models, see e.g. the various studies \,\cite{ApJL2015,GomSolBas2015,ApJ2017,MPLA2017,PLB2017,MNRAS2018a,MNRAS-EPL2018,GBZhao2017,GBZhao2018} and also \cite{Salvatelli2014,Li2016-2015,Murgia2016,Melchiorri2017},
it is remarkable that they can be mimicked by BD-gravity, as we shall show, as this would suggest the possibility that the underlying fundamental theory of gravity might actually be BD rather than GR.


\section{Brans-Dicke gravity}
The Brans-Dicke theory\,\cite{BD} contains an additional gravitational degree of freedom as compared to GR, and therefore it genuinely departs from GR in a fundamental way. The new \textit{d.o.f.}  is represented by the scalar BD field $\psi$, which is nonminimally coupled to curvature, $R$.
The original BD-action reads as follows\,\footnote{In what follows we use  metric and curvature conventions as e.g. in \cite{DEBook,LiddleLyth2000}.}:
\begin{eqnarray}
S_{\rm BD}=\int d^{4}x\sqrt{-g}\left[\frac{1}{16\pi}\left(R\psi-\frac{\oD}{\psi}g^{\mu\nu}\partial_{\nu}\psi\partial_{\mu}\psi\right)-\rL\right]+\int d^{4}x\sqrt{-g}\,{\cal L}_m(\phi_i,g_{\mu\nu})\,. \label{eq:BDaction}
\end{eqnarray}
The (dimensionless) factor  in front of the kinetic term of $\psi$, i.e. $\oD$, will be referred to as the BD-parameter.
The last term of (\ref{eq:BDaction}) stands for the matter action $S_{m}$, which is constructed from the Lagrangian density of the matter fields, collectively denoted as $\phi_i$. There is no potential for the BD-field $\psi$ in the original BD-theory, but we admit the presence of a CC term associated to $\rL$. The dynamics of $\psi$ is such that $\psi(t_0)=1/G\equiv M_P^2$ at present ($t=t_0$), where $G$ is the current Newtonian coupling and $M_P\simeq 1.2\times 10^{19}$ GeV is the Planck mass. Therefore, $\psi$ has dimension $2$ in natural units (i.e. mass dimension squared), in contrast to the dimension $1$ of ordinary scalar fields. The effective value of $G$ at any time is thus given by $1/\psi$, and of course $\psi$ must be evolving very slowly with time. The field equations of motion ensue after performing variation with respect to both the metric and the scalar field $\psi$.  While the  first variation yields
\begin{equation}\label{eq:BDFieldEquation1}
\psi\,G_{\mu\nu}+\left(\Box\psi +\frac{\oD}{2\psi}\left(\nabla\psi\right)^2\right)\,g_{\mu\nu}-\nabla_{\mu}\nabla_{\nu}\psi-\frac{\oD}{\psi}\nabla_{\mu}\psi\nabla_{\nu}\psi=8\pi\left(\,T_{\mu\nu}-g_{\mu\nu}\rL\right)\,,
\end{equation}
the second variation gives the wave equation for $\psi$, which  depends on the curvature scalar $R$. The latter can be eliminated upon tracing over the previous equation, what leads to a most compact result:
\begin{equation}\label{eq:BDFieldEquation2}
\Box\psi=\frac{8\pi}{2\oD+3}\,\left(T-4\rL\right)\,.
\end{equation}
Here we have assumed that both $\oD$ and $\rL$ are constants. To simplify the notation, we have written  $(\nabla\psi)^2\equiv g^{\mu\nu}\nabla_{\mu}\psi\nabla_{\nu}\psi$.
In the first field equation, $G_{\mu\nu}=R_{\mu\nu}-(1/2)Rg_{\mu\nu}$ is the Einstein tensor, and  on its r.h.s. $T_{\mu \nu}=-(2/\sqrt{-g})\delta S_{m}/\delta g^{\mu\nu}$ is the energy-momentum tensor from matter. In the field equation for $\psi$,  $T\equiv T^{\mu}_{\mu}$ is the trace of the energy-momentum tensor for the matter part (relativistic and nonrelativistic). The total energy-momentum tensor as written on the \textit{r.h.s.} of (\ref{eq:BDFieldEquation1})  is the sum of the matter and vacuum parts and adopts the perfect fluid form:
\begin{equation} \label{eq:EMT}
\tilde{T}_{\mu\nu} =T_{\mu\nu}  -\rho_\CC g_{\mu\nu}=p\,g_{\mu\nu}+(\rho + p)U_\mu{U_\nu}\,,
\end{equation}
with $\rho \equiv \rho_m + \rho_r + \rho_\Lambda$ and $p \equiv p_m + p_r + p_\Lambda=p_r + p_\Lambda$,
where $p_m=0$, $p_r=(1/3)\rho_r$ and $p_\Lambda=-\rL$ stand for the pressures of dust matter (which includes baryons and DM), radiation and vacuum, respectively.

As in GR, we have included a possible CC term or constant vacuum energy density, $\rL$, in the BD-action (\ref{eq:BDaction}). The quantum matter fields usually induce an additional, and very large, contribution to $\rL$. This is of course the origin of the CC Problem\, \cite{Weinberg2000,JSPRev2013}\footnote{A recent proposal to alleviate the CC Problem within BD-gravity was made in \cite{GRF2018}.}.

Let us write down the field equations in the flat FLRW metric,  $ds^2=-dt^2 + a^2\delta_{ij}dx^idx^j$. Using the total density $\rho$ and pressure $p$ as indicated above, Eq.\,(\ref{eq:BDFieldEquation1}) renders the two independent equations
\begin{equation}
3H^2 + 3H\fracdpsipsi -\frac{\wBD}{2}\left(\fracdpsipsi\right)^2 = \frac{8\pi}{\psi}\rho\,,\label{Friedmannequation}
\end{equation}
\begin{equation}
2\dot{H} + 3H^2 + \frac{\ddpsi}{\psi} + 2H\frac{\dpsi}{\psi} + \frac{\wBD}{2}\left(\frac{\dpsi}{\psi}\right)^2 = -\frac{8\pi}{\psi}p\,,\label{pressureequation}
\end{equation}
whereas (\ref{eq:BDFieldEquation2}) yields
\begin{equation}\label{eq:FieldeqPsi}
\ddpsi +3H\dpsi = -\frac{8\pi}{2\wBD +3}(3p-\rho)\,.
\end{equation}
Here dots indicate derivatives with respect to the cosmic time and $H=\dot{a}/a$ is the Hubble rate. For constant $\psi=1/G$, the first two equations reduce to the Friedmann and pressure equations of GR, and the third requires $\oD\to\infty$ for consistency. By combining the above equations we expect to find a local covariant conservation law, similar to GR.  This is because there is no interaction between matter and the BD field.  It can indeed be checked by explicit calculation. Although the calculation is more involved than in GR, the final result turns out to be the same:
\begin{equation}\label{eq:FullConservationLaw}
\dot{\rho} + 3H(\rho + p)=\sum_{N} \dot{\rho}_N + 3H(\rho_N + p_N) = 0\,,
\end{equation}
where the sum is over all components, i.e. baryons, dark matter, radiation and vacuum. Usually baryons and radiation are assumed to be conserved, and in some cases one can admit an interaction between DM and vacuum, see e.g. \,\cite{ApJL2015,GomSolBas2015,ApJ2017,MPLA2017,PLB2017,MNRAS2018a,MNRAS-EPL2018} and references therein. Here we take the point of view that all components are separately conserved in the main periods of the cosmic evolution.  In particular, the vacuum component obviously does not contribute in the sum since it is assumed to be constant and $\rho_\CC + p_\CC=0$ .
The expression (\ref{eq:FullConservationLaw}) can also be obtained  upon lengthy but  straightforward computation of the covariant derivative on both sides of Eq.\,(\ref{eq:BDFieldEquation1}) and using the Bianchi identity satisfied by $G_{\mu\nu}$ and the field equation of motion for $\psi$,

\section{Power-law solution}
General analytical solutions to the the system  (\ref{Friedmannequation})-(\ref{eq:FieldeqPsi})  are not known.  However, physical sense guides us into searching for possible solutions in which the BD-field $\psi$ evolves very slowly around the current value $\psi_0=1/G$. Let us seek solutions in power-law form
\begin{equation}
\psi(a) = \psi_0\,a^{-\epsilon}\  \qquad (|\epsilon|\ll1)\,,\label{powerlaw}
\end{equation}
where $a_0=1$ is our normalization for the scale factor at present. The evolution of the effective gravitational coupling is given by $G(a)=1/\psi(a)$.  Obviously $\epsilon$ must be a very small parameter in absolute value since $G(a)$ cannot depart too much from $G$.  For $\epsilon>0$, the effective  coupling increases with the expansion and hence is asymptotically free since  $G(a)$  is smaller in the past, which is the epoch when the
Hubble rate (with natural dimension of energy) is bigger. For  $\epsilon<0$, instead, $G(a)$  would decrease with the expansion. We shall see soon that parameter $\epsilon$ is closely related with $\oD$. \newtext{Let us note that one could also search for solutions of the form $\psi(H)\sim H^{\epsilon}$, or even with the more appropriate normalization $\psi(H)=\psi(H_0)(H/H_0)^\epsilon$,  in which the effective $G(H)\simeq G_0\left(1-\epsilon\ln(H/H_0)\right)$ is once more asymptotically free for $\epsilon>0$. This alternative ansatz should lead to similar solutions as with (\ref{powerlaw}), but here for definiteness we concentrate on the latter. See e.g. \,\cite{FritzschSola1,FritzschSola2} for  studies of variable $G=G(H)$  within dynamical vacuum models. }

From the power-law ansatz (\ref{powerlaw}) we find
\begin{equation}
 \fracdpsipsi = -\epsilon {H}\,,\ \ \ \ \ \ \ \ \ \ \ \
 \fracddpsipsi = -\epsilon\dot{H} + \epsilon^2{H^2}. \label{derivatives}
\end{equation}
Plugging these  relations into the system of equations  (\ref{Friedmannequation})-(\ref{eq:FieldeqPsi}) we encounter, respectively,  the simpler expressions
\begin{equation}\label{epsilonEq1}
H^2 = \frac{8\pi}{3}\frac{1}{\beta\psi}\left(\rho_m+\rho_r+\rL\right)\,,
\end{equation}
\begin{equation}\label{epsilonEq2}
\dot{H}(2 - \epsilon) + H^2\left(3 - 2\epsilon + \epsilon^2 + \frac{1}{2}\epsilon^2\wBD\right) = -\frac{8\pi}{\psi}\left(p_r +p_\CC\right)= -\frac{8\pi}{\psi}\left(\frac{\rho_r}{3} -\rL\right)
\end{equation}
and
\begin{equation}\label{epsilonEq3}
\epsilon\dot{H} + (3\epsilon - \epsilon^2)H^2  = -\frac{8\pi}{2\wBD +3}\frac{1}{\psi}(\rho_m + 4\rho_\Lambda)\,.
\end{equation}
We should recall that $\rho_m=\rho_b+\rho_{dm}$ includes the contribution both from baryons and DM. In the above Eq.\,(\ref{epsilonEq1}) we have introduced parameter $\beta$ as a shorthand for
\begin{equation}
\beta = 1 - \nu_{\rm eff}\,, \qquad \ \ \nu_{\rm eff}\equiv\epsilon\left(1+\frac16\,\oD\epsilon\right)\,, \label{betanu}
\end{equation}
where the small parameter $\nu_{\rm eff}$ represents the tiny deviation of $\beta$ from one. \newtext{The meaning of $\nu_{\rm eff}$ is further commented in the next section.} Both terms in $\nu_{\rm eff}$ can be of the same order, as we shall comment below.

Eq.\,(\ref{epsilonEq1}) takes on a form very close to Friedmann's equation in GR, except that parameter $\beta$ is not exactly equal to $1$ and $\psi$ is not exactly equal to $1/G$. We should remark that the two correction terms in (\ref{betanu}), namely $\epsilon$ and $\oD\epsilon^2$, can be of the same order. In other words, the product $\oD\epsilon$ can be of order one in absolute value. This is because $\oD$ is expected to be large since $\oD\to\infty$ is usually associated to the GR-limit of BD-gravity\,\footnote{ In \cite{Avilez}, for instance, the BD parameter $\oD$ is constrained to satisfy $\oD >890$ at the $99\%$ confidence level on the basis of cosmological data (essentially on CMB). At the much lower scale of the Solar System, the Cassini mission provided data on the parameters of the PPN formalism to infer tighter bounds \,\cite{Will2006}.
For other bounds  on $\oD$ on magnitude and sign, see e.g. \cite{Li13}.  Let us note that,  in general,  the cosmological versus astrophysical bounds are usually treated independently, as they concern very different spatial and temporal scales. In fact, the cosmological constraints can be interpreted in a more general setting, in which BD-gravity is included as a particular case\,\cite{Horndeski74}.}.  Notwithstanding, within the context of the power-law ansatz (\ref{powerlaw}), it is not necessary true that the GR-limit  can be met by just making $\oD$ sufficiently large; we expect that a small value of $\epsilon$ must also be  involved. That this is so is already evident from  Eq.\,(\ref{epsilonEq3}).  While the precise correlation between $\epsilon$ and  $\wBD$ will be elucidated shortly, we will assume the following natural hierarchy for the mixed and higher order terms:
\begin{equation}\label{eq:hierarchy}
|\wBD\epsilon|\sim 1 \gg|\epsilon| \gg |\epsilon^2| \,, \qquad |\wBD\epsilon^2| \sim |\epsilon|\,.
\end{equation}
Let us define as usual the current cosmological mass parameters for each species, $\Omega_N^0=\rho_N^0/\rho_c^0$, where $\rco=3H_0^2/(8\pi G)$ is the critical density today.  Taking into account the separate conservation of the various components,  we can  rewrite (\ref{epsilonEq1}) in the following way:
\begin{equation}\label{eq:E2}
E^2(a) = \frac{a^\epsilon}{\beta}\left[\OMo{a^{-3}} + \ORo{a^{-4}} + \OLo\right]\,,
\end{equation}
where $E=H/H_0$ is the normalized Hubble rate with respect to the current value, and we have implemented the boundary condition $E(a=1)=1$.  Notice that the cosmic sum rule satisfied by the various $\Omega_N$, generalizes in this case as follows:
\begin{equation}\label{eq:SumRule}
\OMo +\ORo +\OLo = \beta=1-\nu_{\rm eff}\,.
\end{equation}
Only for $\epsilon=0$ we have $\nu_{\rm eff}=0$ and then  we recover the usual result $\beta=1$.  From Eq.\,(\ref{epsilonEq1}) it was possible to obtain a first relation among the parameters, expressed by the above sum rule. A second relation can be obtained by combining equations  (\ref{epsilonEq2}) and (\ref{epsilonEq3}). Let us use  the parameter hierarchy (\ref{eq:hierarchy}) to neglect all terms of order $\epsilon$ or higher, which  are much smaller than $|\oD\epsilon|\sim 1$.  We  also multiply the entire Eq.\,(\ref{epsilonEq3}) by $2\oD+3\simeq 2\oD$ (since we assume $|\oD|\gg1$).  The outcome is
\begin{equation}\label{epsilonEq2b}
2\dot{H}+3H^2= -\frac{8\pi}{3}\frac{1}{\psi}\left(\rho_r -3\rL\right)
\end{equation}
and
\begin{equation}\label{epsilonEq3b}
\oD\epsilon\left(2\dot{H} + 6H^2\right)  = -\frac{8\pi}{3}\frac{1}{\psi}(3\rho_m + 12\rho_\Lambda)\,.
\end{equation}
Next we divide these two equations  by $H^2$ and evaluate them  at the present time ($a=1$), hence $\psi\to 1/G$ .  Today's value of the Hubble rate is $H_0$ and we denote by $\dot{H}_0$ the corresponding value of the cosmic time derivative.  Note that radiation can be neglected at present.   So we find
\begin{equation}\label{epsilonEq2c}
2\frac{\dot{H_0}}{H_0^2}+3= 3\OLo
\end{equation}
and
\begin{equation}\label{epsilonEq3b}
\oD\epsilon\left(2\frac{\dot{H}_0}{H_0^2} + 6\right)  = -3\OMo-12\OLo\,,
\end{equation}
where we have used the definition of the cosmological parameters $\Omega_N$  and we have consistently neglected  the ${\cal O}(\epsilon)$ corrections in front of $|\oD\epsilon|\sim 1$. Substituting $\dot{H}_0/H_0^2$ from the first expression into the second, \newtext{and using the cosmic sum rule (\ref{eq:SumRule})}, we arrive at the desired second relation among the parameters:
\begin{equation}\label{eq:omegaDepsilon}
\oD\,\epsilon=-\frac{4-3\OMo}{2-\OMo}+{\cal O}(\epsilon)\,.
\end{equation}
Thus, we have found that the parameter product $\oD\epsilon$ becomes determined in terms of the remaining parameters and hence $|\epsilon|\sim 1/|\oD|$ is small when $|\oD|$ is large, the product of the two parameters being of order one in absolute value. For $\OMo\simeq 0.3$  the mentioned product  is  $\oD\,\epsilon\simeq -1.8 $ and hence $\nu_{\rm eff}\simeq 0.7\epsilon$. This confirms the goodness of the assumed hierarchy of parameters,  Eq.\,(\ref{eq:hierarchy}). However, at this point we still don't know the individual values of $\epsilon$ and $\oD$, except that the first must be small and the second large in absolute value. We can determine these values only after confronting the model with the data, see Sect.\ref{NumericalAnalysis}. The numerical results are displayed for reference in Table 1.

\section{Effective GR-picture of BD-gravity}

We have seen that the Hubble rate within the power-law ansatz (\ref{powerlaw}) is given by (\ref{epsilonEq1}). The latter can be conceived as a sort of Friedmann's equation, but in it the gravitational coupling is actually dynamical: $G(a)=1/\psi(a)=G\,a^\epsilon$. An alternative form for (\ref{epsilonEq1}) is possible, in which $G$ is constant and the dynamics of $\psi$ is effectively transferred to the DE sector. Let us assume that we are in the matter-dominated epoch (MDE) and let $\rho^0_{m}$ be the current matter density. Equation (\ref{epsilonEq1}) takes the form:
\begin{equation}\label{Friedman1}
H^2(a) =\frac{8\pi G}{3}\frac{a^\epsilon}{1-\nu_{\rm eff}}\left(\rho_m^0 a^{-3}+\rL\right)\simeq\frac{8\pi G}{3}\left[\rho_m^0 a^{-3+\epsilon}+\rL a^{\epsilon}+\nu_{\rm eff} a^\epsilon\left(\rho_m^0 a^{-3}+\rL\right)\right]\,.
\end{equation}
where we have expanded linearly in the small parameter $\nu_{\rm eff}$, previously introduced in  (\ref{betanu}). The last term can be cast as\ $\nu_{\rm eff} a^\epsilon(\rho_m^0 a^{-3}+\rL)=[3H^2/(8\pi G)]\,\nu_{\rm eff} (1-\nu_{\rm eff})\simeq {3\nu_{\rm eff} H^2}/(8\pi G)$. Finally, the Hubble rate  can be written in a way that emulates an effective Friedmann's equation with time-evolving cosmological term:
\begin{equation}\label{eq:effective Friedmann}
   H^2=\frac{8\pi G}{3}\left(\rho^0_{m} a^{-3+\epsilon}+\rDE(H)\right)\,.
\end{equation}
\newtext{The piece that plays the role of an effective dynamical dark energy (DDE) density is}
\begin{equation}\label{eq:rLeff}
  \rDE(H)=\rL+\frac{3\,\nu_{\rm eff}}{8\pi G} H^2=\rL+\frac{3\,\nu_{\rm eff}}{8\pi} \, M_P^2H^2\,.
\end{equation}
As we recall, the parameter $\nu_{\rm eff}$ is entirely associated to the dynamics of the BD-field, but here it can be reinterpreted as being responsible for the dynamics of the DE\,\footnote{Worth noticing, the expression (\ref{eq:rLeff}) adopts  the form of the running vacuum model (RVM), see \cite{JSPRev2013,Fossil2008,SolGomRev2015} and references therein, \newtext{where the running parameter is usually denoted $\nu$. We can see that $\nu_{\rm eff}$ in (\ref{eq:rLeff}) mimics the role of the original parameter $\nu$}. The RVM has been shown to be phenomenologically promising to alleviate some of the existing tensions within the $\CC$CDM\,\cite{ApJL2015,GomSolBas2015,ApJ2017,MPLA2017,PLB2017,MNRAS2018a,MNRAS-EPL2018}. \newtext{See also \cite{PRD2009,JCAP2011} for earlier work along these lines}.}.  Concerning the approximations we have made,  notice that $\rL a^{\epsilon}\simeq \rL(1+\epsilon\ln a)$ can be treated just as $\rL$ since its time evolution is negligible as compared to that of  $\nu_{\rm eff}  M_P^2H^2\sim \nu_{\rm eff} \rho_{m}^0a^{-3}$ in (\ref{eq:rLeff}).

For lack of a  better name, we may call such an effective framework, in which BD-gravity is dealt with in an approximate $\CC$CDM form,  the ``GR-picture'' or GR-representation.  Notice that in it the current value of the vacuum energy density is not just given by the original parameter, $\rL$, in the BD-action -- see Eq.\,(\ref{eq:BDaction}) --  but by the quantity   $\rLo\equiv\rDE(H_0)=\rL+{3\nu_{\rm eff}}/{(8\pi}G)\,H_0^2=\rL+\nu_{\rm eff}\,\rho_c^0$.  Given the fact that $|\nu_{\rm eff}|\ll 1$  (since  $|\epsilon|\ll 1$), $\rLo$ is very close to $\rL$.   Furthermore, in the GR-picture the effective matter conservation law involved in (\ref{eq:effective Friedmann}),  that is to say, $\rho_m=\rho_{m}^0 a^{-3+\epsilon}$, is  anomalous. The anomaly is the correction term of order $\epsilon$, namely $\delta\rho_m/\rho_m\simeq \epsilon\,\ln a$.  It triggers  a kind of interaction between matter and the DE in the GR-picture.  In such picture the DE does not appear as the original constant $\rL$, but has acquired a small dynamical component proportional to $H^2$, Eq.\,(\ref{eq:rLeff}). For $\epsilon=0$ (hence $\nu_{\rm eff}=0$) we retrieve, of course, the exact situation of the $\CC$CDM.

Interestingly, the $\sim \oD\epsilon^2$ part of the dynamical term in (\ref{eq:rLeff}) can be thought of as originating from the non-canonical kinetic term of the BD-field, cf. Eq.\,(\ref{eq:BDaction}), interpreted roughly in the manner of an effective potential contribution to the vacuum dynamics:
\begin{equation}\label{eq:noncanonicalKineticE}
 -\frac{\oD}{\psi}g^{\mu\nu}\partial_{\nu}\psi\partial_{\mu}\psi\ \longrightarrow\  \frac{\oD}{\psi}\dot\psi^2\sim \oD \epsilon^2 \psi H^2\sim \oD\epsilon^2 M_P^2 H^2\,.
\end{equation}
The precise and rigorous contribution, however, appears at the level of the field equations and stems from the third term on the \textit{l.h.s} of Eq.\,(\ref{Friedmannequation}). Together with the second term in that equation (which is linear in $\epsilon$ within the context of the power-law solution we are considering) it leads to the running vacuum form (\ref{eq:rLeff})\,\cite{JSPRev2013,SolGomRev2015}.

To complete the  GR-picture  we have to write down the pressure equation as well. The starting point is (\ref{epsilonEq2}). This expression can also be cast  in an approximate $\CC$CDM fashion.  Neglecting once more the small parameter corrections except for those in the dynamical terms, such as $\sim  a^{-3+\epsilon}$ and $\sim \nu_{\rm eff} M_P^2H^2$, and following similar steps as before we can easily obtain the leading expression for the effective acceleration equation in the GR-picture:
\begin{equation}\label{eq:currentacceleration}
\frac{\ddot{a}}{a}=-\frac{4\pi G}{3}\,\left(\rho_m^0 a^{-3+\epsilon}+\rDE(H)+3p_{\Lambda}\right)\,.
\end{equation}
It follows that the equation of state (EoS) for the effective DDE is
\begin{equation}\label{eq:EffEoS}
  w(z)=\frac{p_{\Lambda}}{\rDE(H)}\simeq -1+\frac{3\nu_{\rm eff}}{8\pi \rL}\,M_P^2\,H^2(z)=-1+\frac{\nu_{\rm eff}}{\OLo}\,\frac{H^2(z)}{H_0^2}\,,
\end{equation}
where use has been made of (\ref{eq:rLeff}). Thus, for $\epsilon>0\ (\epsilon<0) $  we have $\nu_{\rm eff}>0\  (\nu_{\rm eff}<0)$ and the effective DDE behaves quintessence (phantom)-like.  For $\epsilon\to 0$ (hence  $\nu_{\rm eff}\to 0$) we have  $w\to -1$ ($\CC$CDM) and only then the BD-parameter is forced $|\oD|\to\infty$ from (\ref{eq:omegaDepsilon}).  From Table 1 off we can read the precise central fitting value $\epsilon= +0.00296$, and hence $\nu_{\rm eff}= 0.696\epsilon=+0.00206$.  It follows from (\ref{eq:EffEoS}) that the current value of the EoS is slightly above the CC divide, namely  $w(z=0)=-0.997$. A plot of $w(z)$  for a redshift range near our time is presented in Fig.\,1. The effective EoS is very close to $-1$ and the DE is therefore of quasivacuum type. Since it approaches $-1$ from above it corresponds to an effective quintessence behavior, which is more pronounced the more we explore the EoS into our past.  This feature  is actually compatible with usual EoS data of  the DE\,\cite{Planck2018}.

It goes without saying that matter is locally and covariantly conserved in BD-theory, as there is no interaction of matter with the BD-field $\psi$ in the fundamental action (\ref{eq:BDaction}). Notwithstanding,  we have shown  here that when we try to encapsulate the observational behavior of BD-gravity  in a strict GR-cliche,  such a theory may effectively appear as interactive quintessence or phantom DE, even though no  quintessence or phantom scalar field is really present. It suggests that the potential monitoring of this sort of ``anomalies'' in the behavior of the $\CC$CDM  could be pointing to BD-gravity as the underlying theory.  This possibility does not exclude that other sources of DE dynamics associated to new fields might also be concomitant, see e.g. \cite{BanerjeePavon2001a}. However, BD-gravity in itself does indeed have the ability to mimic dynamical DE.

Let us also mention that the possibility of observing an anomalous matter conservation law, leading either to a gain or lost in the number density of particles or associated to the slow time evolution of the particle masses with the cosmological expansion, has been proposed in the literature from different points of view,  e.g.  within the context of dynamical vacuum models\,\cite{FritzschSola1,FritzschSola2}.

\begin{figure}
\begin{center}
\label{FigLSS2}
\includegraphics[width=3in, height=2.5in]{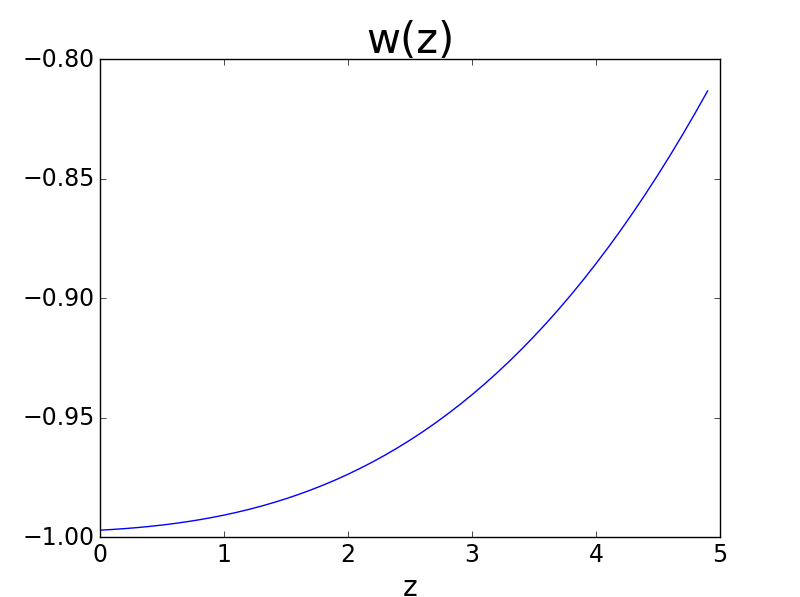}
\caption{\scriptsize Effective EoS for the dynamical dark energy --  see Eq.\, (\ref{eq:EffEoS}) -- as a function of the redshift for a range around our time. We show the EoS curve for the  best fitting value of $\epsilon$ obtained  Table 1, which is positive and therefore it highlights the quintessence regime, as it is obvious from the plot. At present the EoS value is $w(z=0)=-0.997$  and hence very close to the vacuum value from within the quintessence region.  This behavior is perfectly allowed by the observational data and provides an effective quintessence behavior completely unrelated to the existence of fundamental quintessence fields.}
\end{center}
\end{figure}

\section{Structure formation}
The analysis of the global fit requires to consider structure formation as well.  For the models under consideration, we use the standard perturbations equation for the density contrast\,\cite{Peebles1993,LiddleLyth2000},
\begin{equation}\label{eq:DensityContrast}
\delta^{''}_m + \left(3 + a\frac{H^{'}}{H}\right)\frac{\delta^{'}_m}{a} - \frac{4\pi{G}\rho_m}{H^2}\frac{\delta_m}{a^2} = 0\,,
\end{equation}
with, however, the Hubble function corresponding to each one of the models in Table 1 (viz. $\CC$CDM, XCDM and BD). Primes denote differentiation with respect to the scale factor.  For example, in the case of the XCDM the DE density as a function of the scale factor is simply given by $\rho_X(a)=\rho_{X}^0\,a^{-3(1+w_0)}$, with $\rho_{X}^0=\rLo$, where $w_0$ is the (constant) EoS parameter. The corresponding normalized Hubble function is:
\begin{equation}\label{eq:HXCDM}
E^2(a)=\Omega_m^0\,a^{-3}+\Omega_r^0\,a^{-4}+\OLo\,a^{-3(1+w_0)}\,.
\end{equation}
For $w_0=-1$ it boils down to that of the $\CC$CDM with rigid CC term. For $w_0\gtrsim-1$ the XCDM mimics quintessence, whereas for $w_0\lesssim-1$ it mimics phantom DE.
For the perturbation equation we can omit the radiation term since structure formation occurs in the MDE.  For the BD-model the normalized Hubble function is (\ref{eq:E2}), where we can again neglect radiation:
\begin{equation}\label{eq:E2MDE}
E^2(a) = a^{\epsilon}\left[1 + \frac{\OMo}{\beta}(a^{-3} -1)\right].
\end{equation}
 In this case, however, we have to take also into account that   $G(a)=1/\psi(a)=G\,a^\epsilon$ within the context of the power-law ansatz.
The above approximation to structure formation suffices for vacuum models of this sort, as the main effect comes from the change in $H(z)$ as compared to the $\CC$CDM. For models involving additional terms which depend on the vacuum dynamics,  such approximation is also warranted  at subhorizon scales, see \cite{MNRAS2018a} for details. \newtext{We also note that the use of the slow varying  $G(a)$ in Eq.\,(\ref{eq:DensityContrast}) is sufficiently accurate, as shown in previous studies\,\cite{JCAP2011,Elahe2015}.}
Let us now concentrate  on the BD case. The terms in (\ref{eq:DensityContrast}) can be computed  by direct calculation from (\ref{eq:E2MDE}):
\begin{eqnarray}
&& a\frac{H^{'}}{H} =\frac{a}{2}\frac{\left(E^2\right)'}{E^2} =\frac{\epsilon}{2} -\frac{3}{2}\frac{\OMo}{\beta{E^2}}a^{-3+\epsilon}\,,\label{eq:E2MDEb}\\
&&  \frac{4\pi{G(a)}\rho_m(a)}{H^2} = \frac32\frac{\OMo}{E^2}a^{-3+\epsilon}\,.\label{eq:E2MDEc}
\end{eqnarray}


\begin{table*}
\begin{center}
\begin{scriptsize}
\resizebox{1\textwidth}{!}{
\begin{tabular}{ |c|c|c|c|c|c|c|c|c|}
\multicolumn{1}{c}{Model} &  \multicolumn{1}{c}{$H_0$(km/s/Mpc)} & \multicolumn{1}{c}{$\Omega_m^0$}  &\multicolumn{1}{c}{$w_0$} &\multicolumn{1}{c}{$\nu$} &\multicolumn{1}{c}{$\epsilon$}& \multicolumn{1}{c}{$\sigma_8(0)$} & \multicolumn{1}{c}{$\Delta{\rm AIC}$} & \multicolumn{1}{c}{$\Delta{\rm BIC}$}\vspace{0.5mm}
\\\hline
$\Lambda$CDM  & $68.85\pm 0.32$ & $0.298\pm 0.004$ & -1 & -& - & $ 0.797 \pm 0.009$&  - & - \\
\hline
XCDM  & $67.58\pm 0.63$&  $0.308\pm0.006$ & $-0.951\pm0.021$ & - & - & $0.776 \pm 0.012$ &2.91 & 0.89\\
\hline
RVM  & $67.56\pm 0.48$&  $0.305\pm0.005$ & -1 & $0.00131\pm 0.00038$ & - & $0.745\pm 0.018$ & 10.55 & 8.52 \\
\hline
BD  & $67.05\pm 0.61$&  $0.304\pm0.005$ & $\gtrsim-1$& - &$0.00296\pm 0.00088 $ & $0.748 \pm 0.017$ & 9.52 & 7.49\\
\hline
\end{tabular}}

\caption{{\scriptsize Best-fit values for the $\CC$CDM, the generic parametrization XCDM and the power-law solution of the BD-model. \newtext{The corresponding results for the RVM are also included for comparison}. The EoS parameter for the BD case is not constant and its evolution  is displayed in detail in Fig.\,1. Apart from the specific fitting parameters for each model, the remaining parameters are as in the $\CC$CDM
and are not shown.  For convenience we reckon also the values of $\sigma_8(0)$  for each model. The main fitting data are as in \cite{ApJ2017,MPLA2017,PLB2017}.  However, here we have made some updating to the previous sets. For instance, the SNIa data from JLA has been replaced with the Pantheon compilation\,\cite{Scolnic2018}.  We have also updated a few  $f(z)\sigma_8(z)$ data points  from \cite{Shi2018,Okumura2016,Howlett2017,Mohammad2018,GilMarin2018};
 one data point on $H(z=0.47)$ from \cite{Ratsimbazafy2017}, together with BAO data from several sources\,\cite{GilMarin2018,Carter2018,Bourboux2018}; and, finally, the weak-lensing data encoded in the $S_8$-parameter from \cite{Hilderbrandt2017}.  The AIC and BIC parameters are defined in Sect. 6.}}
\end{scriptsize}
\end{center}
\label{tableFit3}
\end{table*}

\begin{table*}
\begin{center}
\begin{scriptsize}
\resizebox{1\textwidth}{!}{
\begin{tabular}{ |c|c|c|c|c|c|c|c|c|}
\multicolumn{1}{c}{Model} &  \multicolumn{1}{c}{$H_0$(km/s/Mpc)} & \multicolumn{1}{c}{$\Omega_m^0$}  &\multicolumn{1}{c}{$w_0$} &\multicolumn{1}{c}{$\nu$} &\multicolumn{1}{c}{$\epsilon$}& \multicolumn{1}{c}{$\sigma_8(0)$} & \multicolumn{1}{c}{$\Delta{\rm AIC}$} & \multicolumn{1}{c}{$\Delta{\rm BIC}$}\vspace{0.5mm}
\\\hline
$\Lambda$CDM  & $69.02\pm 0.31$ & $0.296\pm 0.004$ & -1 & -& - & $ 0.794 \pm 0.009$&  - & - \\
\hline
XCDM  & $68.30\pm 0.60$&  $0.301\pm0.006$ & $-0.973\pm0.020$ & - & - & $0.783\pm 0.013$ & -0.58 & -2.63\\
\hline
RVM  & $68.02\pm 0.46$&  $0.301\pm0.004$ & -1 & $0.00105\pm 0.00037$ & - & $0.751\pm 0.017$ & 6.30 & 4.25 \\
\hline
BD  & $67.82\pm 0.58$&  $0.300\pm0.004$ & $\gtrsim-1$& - &$0.00202\pm 0.00083 $ & $0.758 \pm 0.017$ & 3.78 & 1.73\\
\hline
\end{tabular}}

\caption{{\scriptsize The same as in Table 1 but including now the local measurement of  $H_0$ from \cite{RiessH02018}.}}
\end{scriptsize}
\end{center}
\label{tableFit3}
\end{table*}

In order to solve the above perturbation equation (\ref{eq:DensityContrast})  we have to fix the initial conditions on $\delta_m$ and ${\delta}'_m$  at high redshift, when nonrelativistic matter dominates over radiation and DE, see e.g.\,\cite{PLB2017,MNRAS2018a,MNRAS-EPL2018}.  It is not difficult to see that for $a\to 0$ within the MDE the expressions (\ref{eq:E2MDEb}) and (\ref{eq:E2MDEc}) boil down to constant contributions:  $ a{H^{'}}/{H} \to (\epsilon-3)/{2}$ and ${4\pi{G(a)\rho_m(a)}}/{H^2} \to 3\beta/2$. Under these conditions the differential equation (\ref{eq:DensityContrast}) simplifies to
\begin{equation}\label{diffequationConst}
\delta^{''}_m + \frac12\left(3 + \epsilon\right)\frac{\delta^{'}_m}{a} - \frac{3}{2}\beta\frac{\delta_m}{a^2} = 0\,.
\end{equation}
This equation admits power-law solutions $\delta_m(a) = a^{s}$.  The growing mode can be readily found, and to order $\epsilon$ and $\oD\epsilon^2\sim\epsilon$ we find
\begin{equation}
s =-\frac{1}{4}(1+\epsilon) + \frac{1}{2}\sqrt{\frac{1}{4}(1+\epsilon)^2 + 6\beta}\simeq  1 - \frac{4}{5}\epsilon - \frac{1}{10}\,\wBD\epsilon^2\,.
\end{equation}
Once the initial conditions on $\delta_m\sim a^s$ and ${\delta}'_m\sim s a^{s-1}$  are fixed at high redshift (i.e. small enough $a$) within the MDE, the perturbation equation (\ref{eq:DensityContrast}) with the exact functions (\ref{eq:E2MDEb}) and (\ref{eq:E2MDEc})  can be numerically solved up to our time.

The comparison of the theoretical calculations with the structure formation data in the linear regime is usually implemented with the help of the weighted linear growth $f(z)\sigma_8(z)$, where $f(z)=d\ln{\delta_m}/d\ln{a}$ is the growth factor and $\sigma_8(z)$ is the rms mass fluctuations on $R_8=8\,h^{-1}$ Mpc scales.  The details of its computation were given in previous works and need not be repeated here (see e.g. \cite{MNRAS2018a,MNRAS-EPL2018}).

\begin{figure}
\begin{center}
\label{FigLSS3}
\includegraphics[width=3in, height=2.5in]{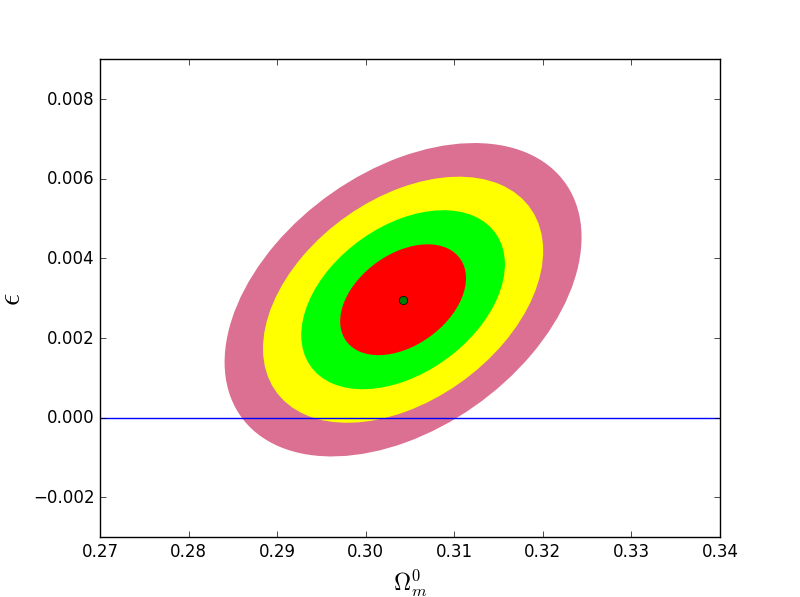}
\caption{\scriptsize Likelihoods contours for the BD model in the ($\Omega^0_m,\epsilon$)-plane for the values $-\ln\mathcal{L}/\mathcal{L}_{\rm max}$ = 2.30,6.18,11.81,19.33 (corresponding to 1$\sigma$, 2$\sigma$, 3$\sigma$ and 4$\sigma$ c.l.) after marginalizing over the rest of the fitting parameters. The central value  corresponds to  $\epsilon=+0.00296$ (cf. Table 1). The $\Lambda$CDM model ($\epsilon=0$) appears comparatively disfavored at roughly  $\sim 3\sigma$ c.l.}
\end{center}
\end{figure}

\section{Numerical analysis and discussion}\label{NumericalAnalysis}

As we have seen, the kind of power-law solutions that we have found allow us to treat the BD-gravity framework in an effective GR form, which we have called the GR-picture of  BD-cosmology.
Let us note that already in the original BD-paper\cite{BD}, power-law solutions were found, but only as powers in the cosmic time (and for zero cosmological constant) rather than powers in the scale factor.  The connection between the two kind of power-law solutions would be possible, in principle, by integrating Eq.\,(\ref{eq:E2}) and finding $t=t(a)$  and then inverting this function to get $a=a(t)$. However, it is difficult in the presence of $\rL\neq 0$ and actually unnecessary.  The direct solutions in terms of the scale factor are more useful and they allow to connect more easily with the observations since these are usually expressed in terms of the redshift: $z=a^{-1}-1$. In the following we put the models under consideration to the test. For the set of observational data used, see Table 1 and references therein.

We perform the statistical analysis of the models in terms of a joint likelihood function, which is the product of the likelihoods for each data source and includes the corresponding covariance matrices,  see e.g.\,\cite{PLB2017,MNRAS2018a,MNRAS-EPL2018} for details.
For a fairer comparison with the $\CC$CDM  we use standard information criteria in which the presence of extra parameters in a given model is conveniently penalized so as to achieve a balanced comparison with the model having less parameters. The  Akaike information criterion (AIC) and the Bayesian information criterion (BIC)  are useful  tools for a fair statistical analysis of this kind. These and other criteria are  discussed e.g. in \,\cite{KassRaftery1995}. The specific definitions of AIC and BIC are
\begin{equation}\label{eq:AICandBIC}
{\rm AIC}=\chi^2_{\rm min}+\frac{2nN}{N-n-1}\,,\ \ \ \ \
{\rm BIC}=\chi^2_{\rm min}+n\,\ln N\,,
\end{equation}
where $n$ is the number of independent fitting parameters and $N$ the number of data points.
The bigger are the (positive) differences $\Delta$AIC and $\Delta$BIC between the $\CC$CDM with respect to the model having smaller values of AIC and BIC (e.g. the BD model) the higher is the evidence in favor of it.
For $\Delta$AIC and $\Delta$BIC in the range  $2-6$ one speaks of ``positive evidence'', and for  $6-10$ one speaks of ``strong evidence'' in favor of the new models. The latter is typically the situation with the  BD-model, see Table 1. In Fig.\,2 we display the corresponding likelihood contours in the plane $(\OMo,\epsilon)$. We can see that  $\epsilon>0$ is favored over $\epsilon=0$ ($\CC$CDM) at roughly $3\sigma$ c.l.  From Table 1 we observe that the XCDM parametrization obtains some positive evidence at $\sim 2\sigma$ c.l. In both cases the quintessence option (i.e.  EoS $w\gtrsim-1$) is consistently preferred. \newtext{For completeness, Table 1 includes also the fitting results for the RVM under the same set of observational data. As expected, the outcome for the  RVM and BD are comparable since the latter mimics  the first. Let us note, however,  that the level of evidence for dynamical vacuum energy has presently decreased with the most recent data as compared to e.g. \cite{ApJ2017}, but it remains still significant ($\sim 3.4\sigma$) and is compatible with the independent studies of \cite{GBZhao2017}.}

The best fitting values $\epsilon\simeq+0.00296$ and $\Omega_m^0\simeq 0.304$  for the BD-model  (cf. Table 1) imply $\oD\,\epsilon\simeq -1.82$  via the constraint (\ref{eq:omegaDepsilon}) and hence $\oD\simeq -615$.
We should emphasize that negative values for $\oD$ are not excluded neither phenomenologically nor theoretically. The limits largely depend
on the parameterization and prior adopted, as well as on the data sets \cite{Avilez,Will2006,Li13}. On theoretical grounds, the negative values can be motivated in the BD effective low-energy models arising from Kaluza-Klein and superstring
theories\,\cite{Kolitch95}. Different applications of the possibility that $\oD<0$  have been considered in the literature, e.g. in \,\cite{BertolamiMartins2000,BanerjeePavon2001b}, sometimes invoking also the presence of a self-interacting potential of the BD-field, or even an interaction between matter and the BD-field\,\cite{Kofinas2015,Kofinas2017}.  In our case the BD framework is much more modest, as we do not introduce any extra field or interaction, we just assume the additional presence of the CC term, as in GR.    

\newtext{In Table 2 we perform the same overall fit to the data, but in this case we include the recent local measurement of $H_0= 73.48 \pm 1.66$ Mpc/s/km from \cite{RiessH02018}, based on Cepheids, in which the tension with Planck  increases to $3.7\sigma$. The aim of this test is to check if the best-fit value of $H_0$ changes significantly as compared to Table 1, where that local measurement was not included as an input. As it can be seen,  the best-fit value for $H_0$  diminishes a bit, but not substantially (viz. $1\%$ or less, depending on the model), and the main conclusions remain. The fit quality of the two models BD and RVM weakens  but they still deliver a better fit to the global observations than the $\CC$CDM, whereas the XCDM now becomes comparable or mildly worse. }

\newtext{It is appropriate to mention here the investigations of\,\cite{Paliathanasis2016}, where  group invariant transformations are used to constrain  flat isotropic and homogeneous cosmological models containing a Brans-Dicke scalar field. They are also used to find out the possible structure of the effective potential\,\cite{Papagianopoulos2017}. The phenomenological impact on the observations, as analyzed in the first work, provides also a competitive fit as compared to the $\CC$CDM, but the level of significance seems to be milder than in our case.  See also \cite{CliftonBarrow2006}  for a framework with energy exchange between the scalar and matter fields in scalar-tensor theories of gravity. }

 \newtext{Let us remark that in our case no effective potential is used for the BD field, and no energy exchange  is produced in the original BD picture, although it is mimicked in the GR picture. It should be worth exploring if such analogy with the RVM can be extended to describe the early universe and inflation, as it is indeed possible in the case of the RVM itself (see \cite{LBS2012,SolGomRev2015}). All the more if we take into account that the RVM can be connected to the action of anomaly-induced inflation\,\cite{Fossil2008}.  This fact and the relationship discussed here with a class of physically meaningful solutions of the BD theory seem to indicate that the RVM encodes the effective behavior of a family of phenomenologically interesting cosmologies.}


\vspace{0.3cm}

\section{Conclusions}
Although  BD-gravity theories  are fundamentally different from GR, we have found that they admit power-law solutions which can be interpreted in GR form. The effective behavior is  $\CC$CDM-like with, however, a mild time-evolving  quasivacuum component.  The corresponding EoS is very close to $-1$ and mimics quintessence.   Our fit to the data shows that such an effective  representation of BD-gravity can be competitive with the concordance model with rigid $\CC$-term. This might help to smooth out some of the tensions afflicting the $\CC$CDM in a way similar to the running vacuum model, see e.g.\,\cite{ApJ2017,MPLA2017,PLB2017,MNRAS2018a,MNRAS-EPL2018}. We conclude that finding  traces of vacuum dynamics, accompanied with apparent deviations from the standard matter conservation law could be the ``smoking gun'' pointing to the possibility that the underlying gravity theory is not GR but BD.

\vspace{1cm}

{\bf Acknowledgements}

We are partially supported by projects  FPA2016-76005-C2-1-P (MINECO), 2017-SGR-929 (Generalitat de Catalunya) and MDM-2014-0369 (ICCUB). JdCP is also supported by a  FPI fellowship associated to the project FPA2016-76005-C2-1-P.

\end{document}